\documentclass[lettersize,journal]{IEEEtran}
\usepackage{amsmath,amsfonts}
\usepackage{algorithmic}
\usepackage{algorithm}
\usepackage{array}
\usepackage[caption=false,font=normalsize,labelfont=sf,textfont=sf]{subfig}
\usepackage{textcomp}
\usepackage{stfloats}
\usepackage{url}
\usepackage{verbatim}
\usepackage{graphicx}
\usepackage{cite}
\usepackage{xcolor}

\usepackage[justification=centering]{caption}

\begin{document}

\title{Large Language Models powered Malicious Traffic Detection: Architecture, Opportunities and Case Study}

\author{
        \IEEEauthorblockN{
        Xinggong Zhang, Haotian Meng, Qingyang Li, Yunpeng Tan, and Lei Zhang
        }\\
        \thanks{This paper was sponsored by the NSFC grant (62431017). We gratefully acknowledge the support of Zhongguancun Laboratory, Key Laboratory of Intelligent Press Media Technology.}
        \thanks{Xinggong Zhang (zhangxg@pku.edu.cn), Qinqyang Li and Yunpeng Tan are with Peking University; Haotian Meng (Co-first authors) is with China Unicom; Lei Zhang (Corresponding author) is with Zhongguancun Lab.}
        }



\maketitle

\begin{abstract}
Malicious traffic detection is a pivotal technology for network security to identify abnormal network traffic and detect network attacks.
Large Language Models (LLMs) are trained on a vast corpus of text, have amassed remarkable capabilities of context-understanding and commonsense knowledge. This has opened up a new door for network attacks detection.
Researchers have already initiated discussions regarding the application of LLMs on specific cyber-security tasks. Unfortunately, there remains a lack of comprehensive analysis on harnessing LLMs for traffic detection, as well as the opportunities and challenges.
In this paper, we focus on unleashing the full potential of Large Language Models (LLMs) in malicious traffic detection. We present a holistic view of the architecture of LLM-powered malicious traffic detection, including the procedures of Pre-training, Fine-tuning, and Detection.  
Especially, by exploring the knowledge and capabilities of LLM, we identify three distinct roles LLM can act in traffic classification: \textit{Classifier, Encoder, and Predictor}. For each of them, the modeling paradigm, opportunities and challenges are elaborated. 
Finally, we present our design on LLM-powered DDoS detection as a case study. The proposed framework attains accurate detection on carpet bombing DDoS by exploiting LLMs' capabilities in contextual mining. The evaluation shows its efficacy, exhibiting a nearly $35$\% improvement compared to existing systems.
\end{abstract}


\section{Introduction}
Malicious traffic detection is the cornerstone of network security since it is about the classification and identification of network attack traffic. It usually oversees network behavior and detects network anomalies that might compromise the security, integrity, and functionality of networks. According to Cloudflare reports~\cite{cloudflare}, each day in 2024 witnessed 209 billion network attacks events, marking a 53\% annual increase. The largest recorded DDoS attack in 2024 even surpassed 5.6 terabits per second in traffic volume.
It remains critical challenges for both academia and industry to detect and prevent network attacks timely and precisely.

With the exponential growth of network traffic and evolving natures of attacks, existing malicious traffic classification methods, such as rule-based, machine-learning or deep-learning and anomaly-based methods, are confronting critical challenges~\cite{meng2023}. \textbf{Rule-based} approach has been extensively embraced by the industry for its practicality and effectiveness. But the hand-crafted rules are time-consuming and require a great deal of expertise. Supervised \textbf{machine-learning}~\cite{shen2020} or \textbf{deep-learning} methods~\cite{Tan2024} require a amount of labeled data for the training, which is impractical due to the labor-intensive labeling work and huge amounts of network traffic. Unsupervised \textbf{anomaly detection} methods~\cite{meng2023} can detect zero-day attacks by identifying the pattern that deviate from normal network traffic, but it faces the issues of traffic drift.     

Recently, \textbf{Large Language Models} (LLMs)~\cite{scaling} have demonstrated powerful potentials in natural language processing and computer vision, such as language understanding, image classification, and multi-modal learning, etc. LLMs, including BERT~\cite{bert}, GPT~\cite{gpt} series, etc. are based on the Transformer neural network architectures, and learn the in-context pattern of token sequence from a vast corpus of text using self/semi-supervised learning techniques, 
The key capabilities of LLMs on \textit{context-understanding} and embedded \textit{vast knowledge} open a new door for network attack detections~\cite{networkLLM}.

Unfortunately, there is still a lack of comprehensive elaboration on the current \textbf{LLM-powered} malicious traffic detections, its overall architecture, as well as the opportunities and challenges therein. A few of surveys~\cite{access2024,dan2025} have summarize the recent researches on the application of LLMs in offensive and defensive cyber-security. Most of them exploit LLMs' capability on nature language processing such as network intent understanding, network operation, security knowledge Q\&A, etc. Few of them explore LLMs in traffic classification. They don't delve into the technical architecture and illustrating a clear picture of how LLMs enable malicious traffic detection.

In this paper, we fill this gap from the perspective of LLMs' capability. We present an architecture of LLM-powered malicious traffic detection. Especially, by exploring the potential of LLM, we introduce three roles of LLMs can act in traffic classification, \textbf{Classifier, Encoder, and Predictor}. For each role, the detection paradigm, opportunities and challenges are elaborated. We try to answer the three question. 1) \textit{How to exploit LLMs' potentials in malicious traffic detections}. 2) \textit{How to enable LLMs  to understand non-language network traffic}. 3) \textit{What capability LLMs can provide in the detection}. To the best of our knowledge, this paper is one of the first works that systematically present the architecture and paradigm of LLM-powered malicious traffic detection. 

In summary, this paper makes the following contributions:
\begin{itemize}
    \item We present an holistic view of the architecture of LLM-powered malicious traffic detection, from the perspective of exploring LLMs' potentials.
    \item We point out the roles LLM can play in traffic classification, and corresponding opportunities and challenges.
    \item We present our design as a case study where LLMs act as an Encoder in DDoS detection. 
\end{itemize}

The rest of the article is organized as follows: Sec.~\ref{Sec-arch} present a holistic view of the architecture of LLM-powered malicious traffic detection. In Sec.~\ref{Sec-oppo}, the LLM's roles, opportunities and challenge are introduced in details. Sec.~\ref{Sec-case} demonstrates a case study of LLM-power DDoS detection. Finally, the conclusion and future work are given in Sec.~\ref{Sec-conclu}.

\section{The Architecture of LLM-powered Malicious Traffic Detection}
\label{Sec-arch}
\begin{figure*}
    \centering
    \includegraphics[width=1\linewidth]{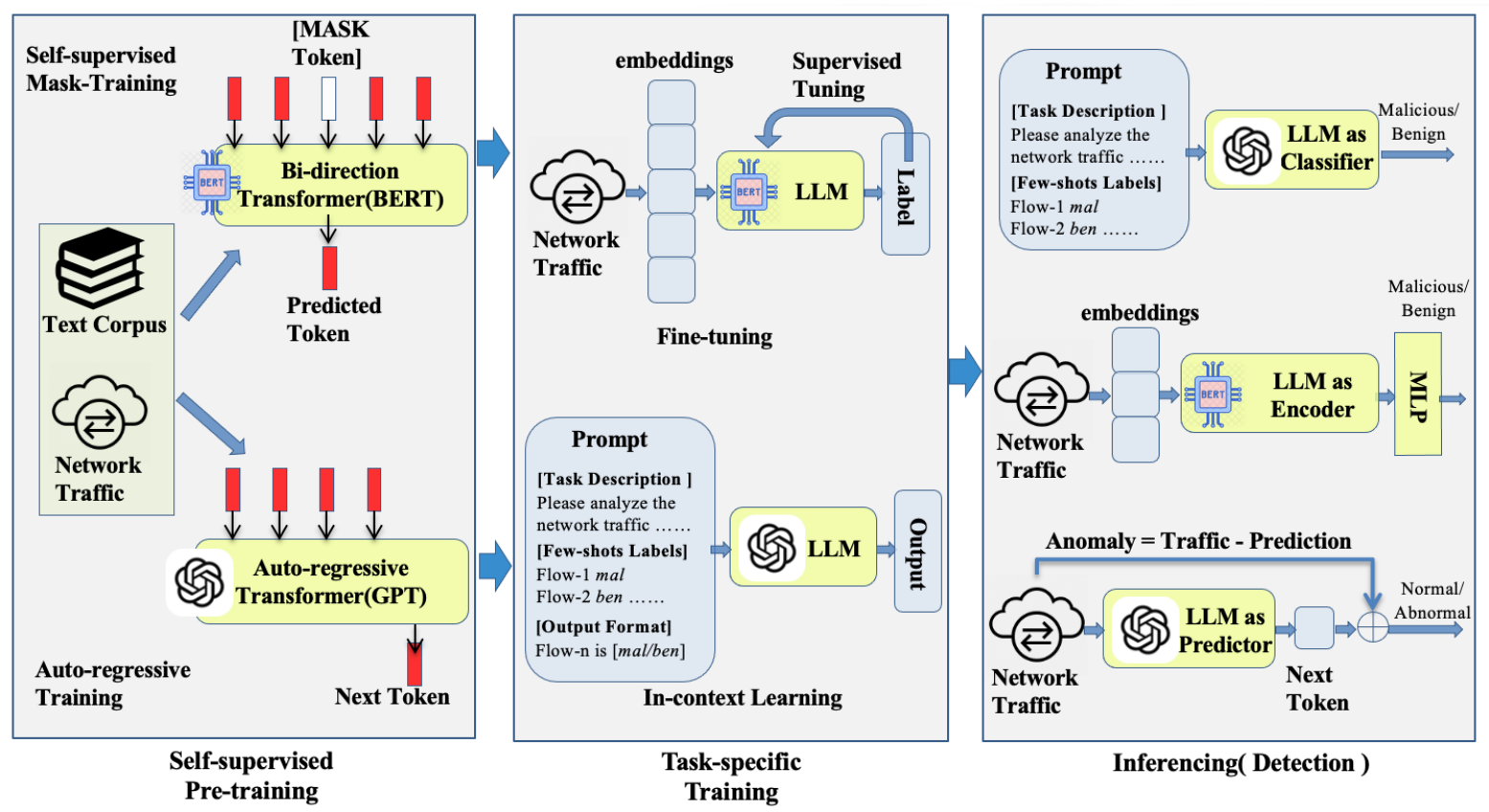}
    \caption{Architecture of LLM-powered Malicious Traffic Detection.}
    \label{fig:image_arch}
\end{figure*}

In this section, we introduce the architecture of LLM-powered malicious traffic detection. As in Fig.~\ref{fig:image_arch}, the framework comprises three sequential stages, i.e., \textbf{Pre-training}, \textbf{Task Training}, and \textbf{Detection}.   

\subsection{Self-supervised Pre-training}
LLMs pre-training is a crucial technology of training a model with transfer learning capability. It is a self-supervised learning method, which means the model learns from the data itself without relying on explicit human-labeled supervision. The aim is to enable the model to learn the general rules and semantic knowledge of the sequential data, such as the relationships between data, the structure and semantics of sequences, etc. The pre-trained models can be fine-tuned in various downstream tasks with improved performance and reducing the demand for a large amount of labeled data.

\textbf{Pre-training Data} for LLM-powered malicious traffic detection could include text corpus and network data. A large amount of text data, such as academic papers, books, web pages, etc. allow the model to learn general language patterns and vast knowledge.
Network data include packet-level, flow-level and topology-level network traffic data which make LLM learn network communication pattern.

\textbf{Pre-trained Models}, also called foundational model, can be trained with network specific data. But recent studies~\cite{scaling} have revealed that the model trained on nature-language text is able to understand multi-model data. Thus, LLM foundation model pre-trained on text data will be a proper choice for malicious traffic detection.

{Pre-training Methods} can be primarily categorized into two paradigms: \textbf{Masked Training} and \textbf{Auto-regressive Training}, each employing distinct learning mechanisms to enable model generalization:

\textbf{Masked Training:} Masked pre-training involves training models to predict masked tokens in sequences. The canonical model is developed by Google, Bidirectional Encoder Representations from Transformers(BERT)~\cite{bert}. The model is presented with a large amount of data where certain tokens are masked. The LLM's task is to predict these masked tokens based on the surrounding context. This training helps the LLM to understand context within data. 

\textbf{Auto-regressive Training:} The model is trained to predict the next token in a sequence given the previous tokens. The well-known model of this type is Generative Pre-trained Transformer(GPT)~\cite{gpt}, a series of pre-trained language models developed by OpenAI.  
The model processes sequences in a left-to-right manner, using a causal attention mechanism. This enables it to learn the sequential patterns of network traffic in malicious traffic detection.

\subsection{Task-specific Post-training}
Task-specific post-training is a crucial step in adapting a pre-trained foundation model to a specific task, such as traffic classification, anomaly detection, etc. It refines the general knowledge acquired during pre-training, enabling LLM to develop task-specific capabilities for malicious traffic detection. Specifically, task-specific post-training can be categorized into two primary paradigms: In-Context Learning and Fine-tuning.

\textbf{In-Context Learning:} In-Context Learning (ICL)~\cite{prompt} allows the LLM to learn from the examples provided within the input context, Prompt, without explicitly updating the model's parameters. In the context of malicious traffic detection, this paradigm enables zero-shot/few-shot inference with a prompt consisting of a description of a network security scenario, along with some examples of benign and malicious activities.

ICL is powerful as it enables the model to quickly adapt to new tasks without requiring extensive re-training LLM models, making it highly efficient for addressing time-evolving network attacks. But it falls short of detection accuracy.


\textbf{Fine-tuning:}~Task Fine-tuning focuses on training the LLM for a specific task by supervised learning with labeled training data, as shown in Fig.~\ref{fig:image_arch}. The pre-trained model are used as an initialization whose parameters are updated by supervised gradient. This significantly reduces the amount of data and time required compared to training models from scratch.

The advantage of this approach~\cite{access2024} is to optimize the model's performance with high accuracy. But the drawback is that a lot of labeled data for training is needed, and updating model parameter is computation-consuming.

\subsection{Detection}
According to the potential capability of LLM, it can perform three crucial roles in malicious traffic detection: Classifier, Encoder, and Predictor. Each role contributes uniquely to enhancing the effectiveness and efficiency.


\textbf{Classifier:}~Exploiting the commonsense knowledge of LLM, it is possible for LLM to serve as an intelligent Agent~\cite{gpt} in malicious traffic detection.
As in Fig.~\ref{fig:image_arch}, The LLM is fed into Prompt, a text description about task introduction, some examples of network flow description, and your questions. The LLMs answer whether the flow is benign or malicious.
It applies the knowledge it has learned during pre-training and few-shots to make a classification decision.

\textbf{Encoder:}~Exploiting the capability of context understanding, LLM often act as a feature extractor~\cite{encoder2024access}, i.e., representation learner.
As an encoder, the LLM extracts meaningful representations or features from network traffic, often in the form of a vector. This encoded representation captures the essential features and relationships within the network traffic.

Comparing to the traditional method of human feature engineering, the encoding process by the LLM helps in condensing complex traffic data into a format that can be easily processed, facilitating more efficient threat detection.

\textbf{Predictor:}~By exploiting the capability of LLMs on masked-token prediction or next-token auto-regressively generation~\cite{prediction}, LLM can predict future traffic pattern. As a predictor, the LLM aims to forecast future activities or trends related to network traffic. It uses historical data, current network status information, and the knowledge it has acquired during training to make predictions.

For example, based on past patterns of network traffic, LLM can predict the future traffic patterns. If the actual traffic deviate from the prediction, an anomaly event is detected.    

\section{Opportunities and Challenges}
\label{Sec-oppo}
In this section, we discuss the roles, research opportunities and challenges in LLM-powered malicious traffic detection in details. 
Specifically, we focus on the novel diagram of LLM as Classifier, Encoder, and Predictor.

\subsection{LLM as Classifier}
\begin{figure}
    \centering
    \includegraphics[width=1\linewidth]{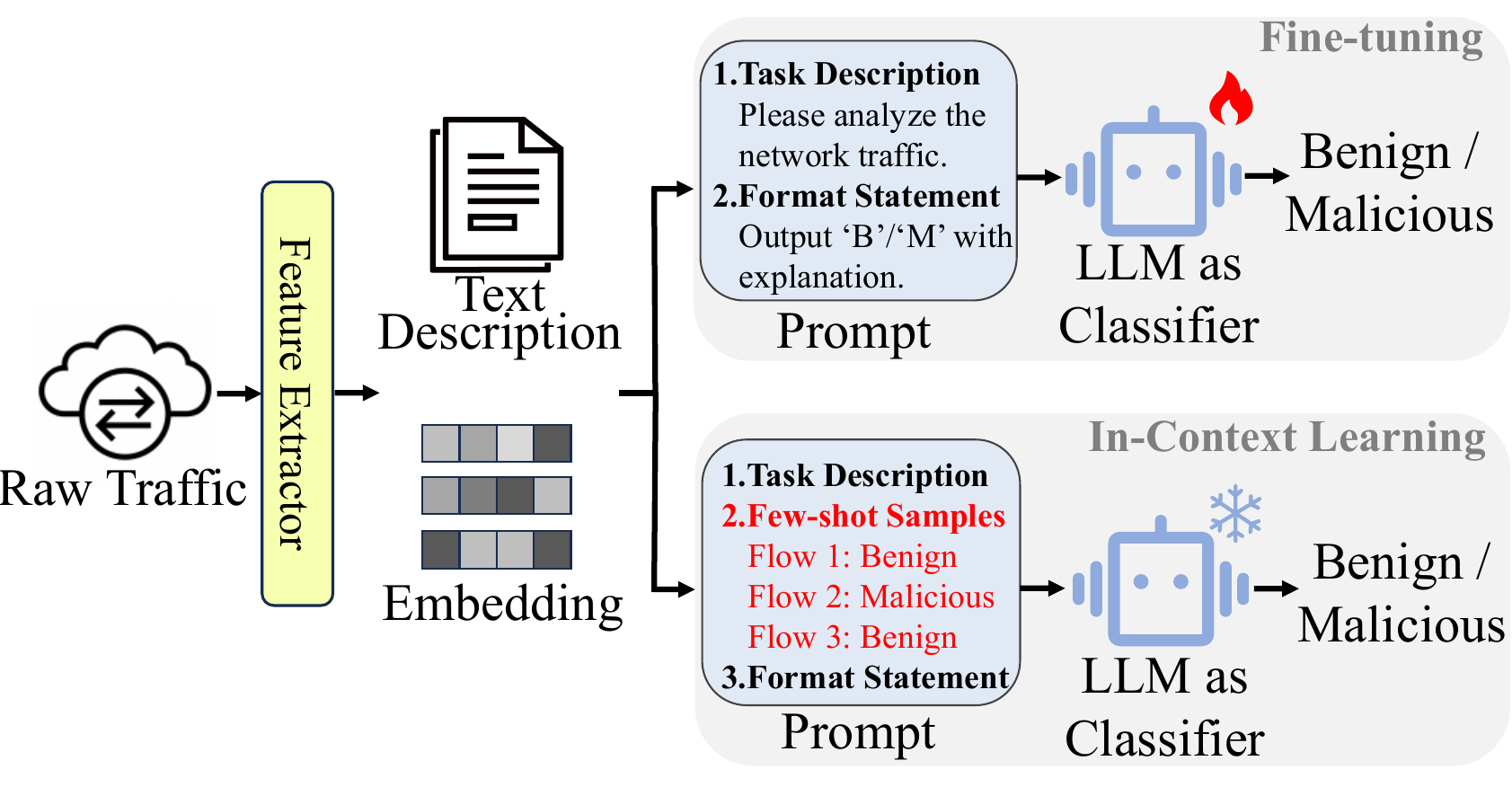}
    \caption{LLM as Classifier}
    \label{fig:image_classifier}
\end{figure}
Inspired by the concept of \textbf{LLM agent}, where an {agent} is trained to handle planning, decision-making, and control, the idea of making an LLM an intelligent agent for malicious traffic detection is promising~\cite{prompt}. The network traffic data and carefully-crafted Prompts are feed into the LLM. It classifies network traffic by the guidance provided in the Prompts, leveraging its pre-trained language-understanding knowledge and pattern-recognition capabilities. The wealth human-knowledge in LLM provides great opportunities for malicious traffic detection, but also comes with challenges.     

As showed in Fig.~\ref{fig:image_classifier}, the main components of LLM-based classifier consists of three crucial modules: Traffic Processing, Prompt Construction, and LLM instruction fine-tuning or ICL. The modules work together to make LLM understand network traffic and inference which flow is malicious.

\textbf{Traffic Processing.} Before feeding into the LLM, raw network traffic first needs to be processed to a form that the LLM can understand. There are two main processing methods: text description and embedding learning.

\textit{Text Description:} Text description involves converting raw network traffic data into language text. For example, network packet headers, which contain details like source and destination IP addresses, port numbers, and protocol types, can be translated into texts. This text-based representation allows the LLM to utilize its natural-language-processing capabilities to analyze the network data. But it has limitations in handling numerical data with high precision, since text could not capture complex patterns in numerical data.

\textit{Embedding Learning:} Embedding learning aims to transform the raw network traffic data into a numerical representation, often in the form of vectors or tokens. For network traffic data, algorithms can be used to extract features such as the frequency of different protocol types, the distribution of packet sizes, and the relationship between source and destination IP addresses. These features are then mapped into a vector space. This numerical representation can be more easily processed by the LLM, enabling it to perform operations such as similarity comparison and pattern recognition more efficiently.

\textbf{Prompt.} The prompt, which is intrinsically hand-crafted text question, plays a crucial role in guiding the LLM to produce accurate attack-detection results. Prompt engineering is still an open topic how to design good Prompt to better explore the potential of LLM's extensive human knowledge.  

Here is an example of Prompt template. It includes: 

\textit{Task Description:} This part clearly defines the objective and instruction of the network threat detection task. By providing a clear task description, the LLM is better able to focus its analysis on the relevant aspects of the network traffic.

\textit{Few-shot Samples:} Usually there are a small number of labeled examples to illustrate the input data and their corresponding traffic classifications. For instance, a few-shot sample could include an example of benign network traffic with a description of why it is considered benign. These samples help the LLM learn the patterns and features associated with the Malicious/Benign labels. They are equivalent to the training data set in supervised learning.    

\textit{Output Statement:} It specifies the expected output format of the LLM detection. For example, \textit{Please respond with either 'Benign' or 'Malicious' to indicate the nature of the network flows, and provide a brief explanation for your classification}. This ensures that the LLM's response is in a consistent format without nonsensical generated text.

\textbf{ICL or Fine-tuning.} LLM should be fine-tuned to know how to detect malicious traffic as mentioned in~\ref{Sec-arch}. Two types of task-specific model adaptation are both applicable to traffic detection: Fine-tuning and In-Context Learning. 

Instruction fine-tuning is a supervised learning which needs a few of labeled texture samples. 
As the results validated in~\cite{prompt}, it can achieve higher precisions. But it requires modifying the LLM's parameters through supervised gradient. This will allow the LLM to better adapt to the specific requirements of network threat detection tasks.

Instead, in-context learning construct a few instances of known traffic patterns and benign patterns in Prompt. The LLM then uses only few samples to make inferences about new, unseen network traffic data. This approach is more flexible as it can quickly adapt to new attacks without the need for parameter updates. But the detection precision is lower because the in-context samples may have distribution bias.

Although LLM classifiers exhibit amazing potentials in malicious traffic detection, there are still many non-intuitive key challenges that need to be overcome.

\textit{Bias of Samples Distribution:} Whether in fine-tuning or in-context-learning, the labeled samples should be provided. When the distribution of the training dataset diverges from that of the testing dataset, it invariably leads to a reduction in detection accuracy.

\textit{Hallucination:} LLMs are probability models of language tokens, prone to hallucination anyway. This may generate false positive results in the context of malicious traffic detection.
For example, the LLM might misinterpret benign network traffic fluctuations as signs of a malicious attack. This could be due to its over-reliance on certain patterns in the training data.

\textit{Computation Consumption:} Malicious traffic detection often demands real-time or near-real-time analysis of traffic. However, due to the computation-consumption of LLM, it may takes LLM to inference with unacceptable latency.

\subsection{LLM as Encoder}
\begin{figure}
    \centering
    \includegraphics[width=1\linewidth]{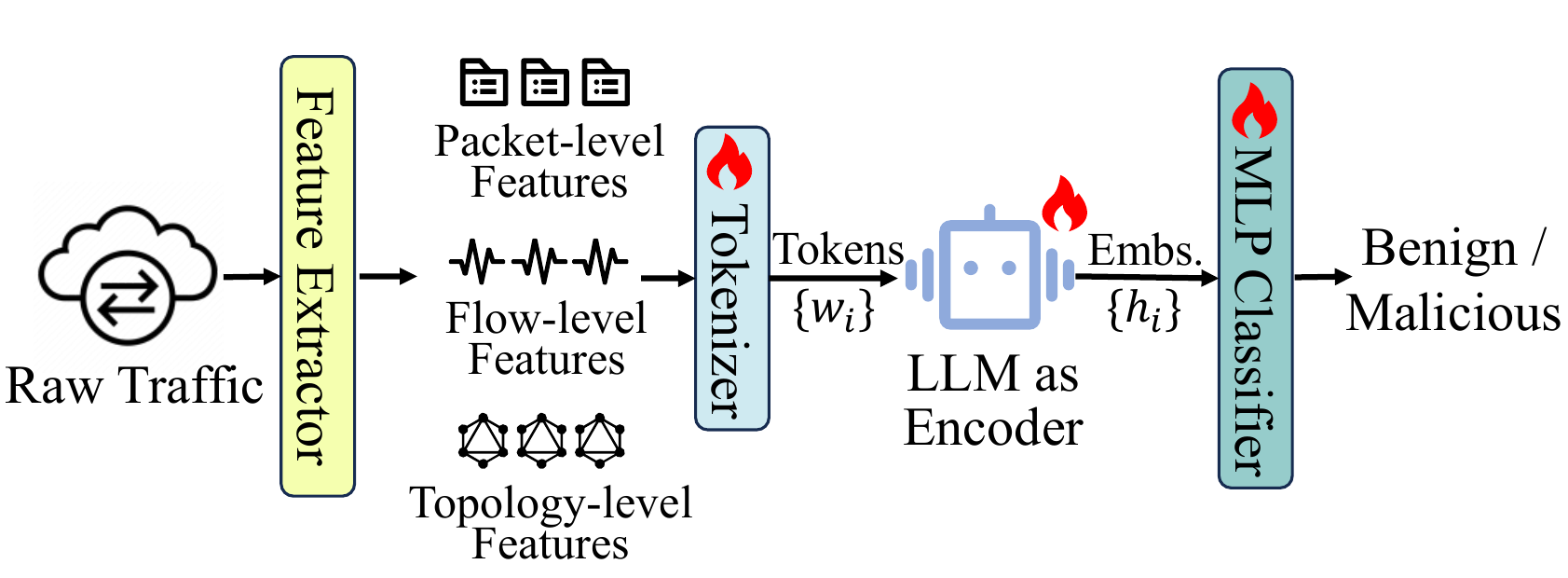}
    \caption{LLM as Encoder}
    \label{fig:image_encoder}
\end{figure}
In malicious traffic detection, LLM can also act as a \textbf{representation learner}, the Encoder. It transforms network raw traffic into a vector representation, i.e., tokens or embeddings. This vector encapsulates the essential characteristics of the network data, making it easier to process and classify further. LLMs can not only learn explicit features but also mine latent features that humans cannot understand, such as in-context correlations of network traffic data.

As shown in Fig.~\ref{fig:image_encoder}, the typical pipeline of LLM as Encoder based network threat detection consists of three parts: Tokenizer, LLM Encoder, and MLP classifier.

\textbf{Tokenizer:} Raw network traffic data are extracted into explicit features, such as packet-level, flow-level and topology-level statistical features. These features are  broken down into smaller units, called tokens. This tokenization process is crucial as it converts the complex and unstructured network traffic data into a format that can be processed by LLM. Traffic can be organized into sequences by temporal or spacial position. The position of token in the sequence has significant effects on the extracted features. 

\textbf{LLM Encoder:} The tokens are fed into the LLM Encoder. The LLM Encoder, based on the pre-trained LLM architecture, processes the tokens. It uses the learned patterns and knowledge from its pre-training to generate a vector representation for the input tokens. Usually two types of LLM model can be used in malicious traffic detection, BERT of bidirectional encoder, and GPT of auto-regressive embeddings generation.

Given a sequence of input tokens $w_i$, the encoded representations can be presented as:
\begin{equation}
    h_1, h_2, ..., h_T = Encoder(w_1,...,w_T)
\end{equation}
Here $h_i$ is the output embeddings which is used for classification in the following steps.

\textbf{MLP classifier:} By adding a output layer to a pretrained LLM, traffic classification results can be obtained. Usually a simple MLP (Multi-Layer Perceptron) is used as the output layer. The MLP classifier takes the vector representation output by the LLM Encoder and uses it to classify the network traffic as benign or malicious.

Although LLM as encoder could provide great potential to mine data features instead of handcrafted feature engineering. But it also exists some challenges to be solved.

\textit{Data Sequentialization:} Network traffic data is not always sequential in nature. For example, packets are sequential as arriving in a specific order over time. But flows in a network there isn't temporal sequential feature. It is necessary to sequence data before tokenizing them.

\textit{Generalization:} An LLM Encoder trained or fine-tuned on a specific set of network traffic may not be able to effectively generate accurate vector representations for networks with different traffic pattern or attack vectors. It's a critical challenge for LLM encoder to extract features against traffic drifts.    

\textit{Scalability:} As the scale of network traffic and the number of network devices continue to grow exponentially, the scalability of LLM-based encoders becomes a critical issue. Scaling up the LLM encoder to handle this massive data volume requires not only more computational resources but also efficient distributed computing.

\subsection{LLM as Predictor}
\begin{figure}
    \centering
    \includegraphics[width=1\linewidth]{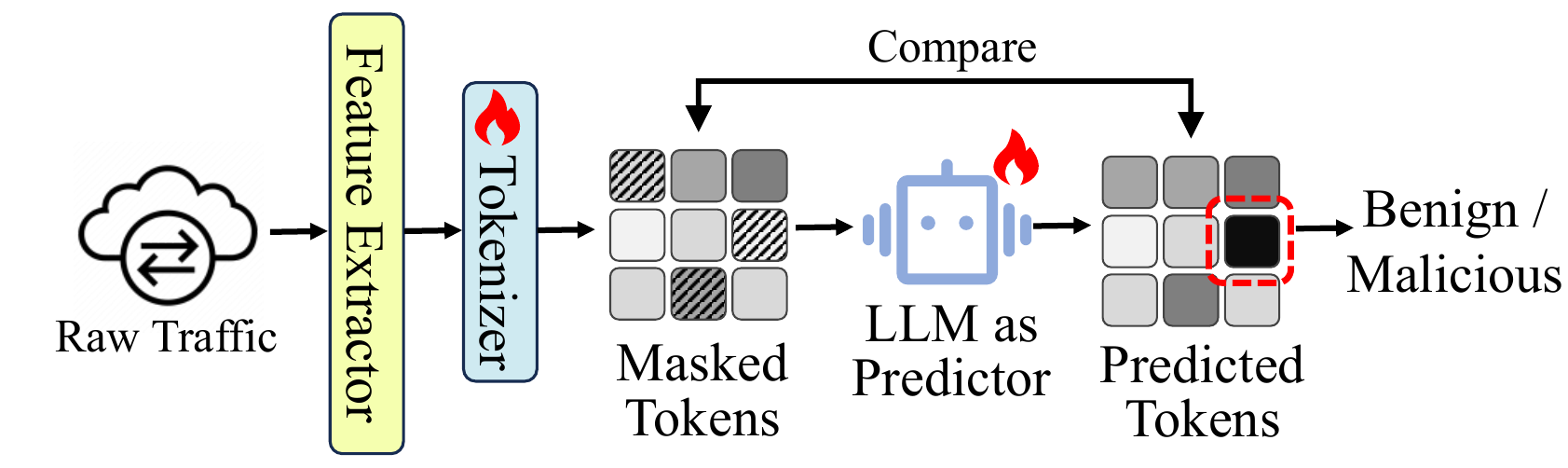}
    \caption{LLM as Predictor}
    \label{fig:image_predictor}
\end{figure}
As the classical anomaly detection methods, LLMs could act as Predictor to detect network traffic anomalies. 
The LLMs, being trained with normal or baseline traffic during its pre-training or fine-tuning phase, predict the most likely next or masked token in a network traffic sequence. 
If the actual traffic deviates from the predicted tokens, it could indicate a anomaly.

Fig.~\ref{fig:image_predictor} demonstrate the pipeline of LLM-as-Predictor. The language model of BERT and GPT can be used to execute anomaly detection. For BERT model, the masked token is predicted by bi-direction contexture sequence. For GPT model, the next token in the sequence is predicted in a auto-regressive way. LLM is trained or fine-tuned only with benign traffic data. It is expected to exhibit significant errors and low probability when they encounter abnormal traffic during the test.

As in Fig.~\ref{fig:image_predictor}, the framework of LLM-based anomaly detection consists of three parts: 

\textbf{Tokenizer:} As network traffic is captured, it tokenizes them and then masks parts of tokens for predictions. For BERT model, the key tokens in the sequence are masked. For GPT model, the tokens are fed into LLM one-by-one.  

\textbf{Fine-tuned LLM:}  Use the normal network traffic data (tokenized and labeled) to fine-tune the LLM. The fine-tuning process adjusts the pre-trained weights of the LLM to better fit the characteristics of the normal network traffic. This involves training the LLM to predict the next token or masked token in a normal traffic sequence.

\textbf{Anomaly Detection:} In the detection, the prediction error is calculated by cross-entropy loss that occurs between the predicted token and the input traffic. Additionally, the predictive probability is a value with the highest probability of token that can appear in the prediction. When the probability of a predicted token is low or the prediction error is large, the respective traffic traffic is considered rare to find in the normal context and is identified as an anomaly. 

LLM-as-predictor explore the LLMs' capability on token generations. It has a crucial assumption that abnormal traffic's \textbf{in-context pattern} is different from that of normal traffic. It faces several challenges:

\textit{Traffic Tokenization:} Converting network traffic into a format suitable for an LLM's token-based prediction is not straightforward. Network traffic data is often in binary or a complex structured format. Translating this data into a sequence of tokens that the LLM can understand while retaining its semantic meaning is a challenge.

\textit{Pattern Drift:} As in classical anomaly-based detection, the distribution of network traffic data changes continuously over time. This requires the model to adapt to these changes in a timely manner. Otherwise, the probability scores calculated by the model based on historical data may not accurately reflect the current abnormal situations, resulting in a large number of false positives or false negatives.

\section{Case Study: A Design of LLM-based DDos Detection}
\label{Sec-case}
In this section, we introduce a system design as a case study that LLMs detect Carpet Bombing DDoS Attacks (CBA) with high accuracy and especially zero-shot capability.

\subsection{LLM-based Encoder in Flow-Correlation Mining}
This system is design to detect a typical network attack, Carpet Bombing DDos Attacks (CBA), on the backbone network of telecom operators~\cite{2024dollm}. CBA is currently an increased threat that targets large-scale network infrastructure such as public cloud, internet data centers, and ISP networks. It simultaneously floods a group of IPs, and overwhelms their shared bottleneck link. 
Since the end-host is not the explicit target but the bottleneck link is, the legacy victim-based detection method is not applicable anymore~\cite{iwqos}.

Leveraging the potential of \textbf{LLM as Encoder}, the proposed LLM-based system mines the multi-flows contextual correlations to enhance the detection features. Since the attackers during a CBA usually use the same attack tools, there exist strong correlations among malicious flows. Hence, we can use LLMs' self-attention mechanism to capture the correlations between tokens (network flows) in sequence. 

The framework consist of four parts: Flow Sequentializer, Flow Tokenizer, LLM Encoder and Classifier. It organizes network flows into a flow sequence by Flow Sequentializer, and projects each flow into the semantic space of LLMs by Flow Tokenizer. LLM acts as encoder to transforming the token into flow embedding. The inter-flow correlations are explored by the LLMs backbone. Finally, the detection task is modeled as a binary classification task, and the flow will be classified based on the flow embeddings outputs by the LLMs.

In the pipeline, it deals with two challenges, flow-sequence construction and modality alignment between flows and text.

\textbf{Flow Sequentializer.} Since the position encoding of LLMs, the position of each flow in a sequence is important. However, due to the huge number of users in the backbone network and the network jitter, the flows will be randomly arrived. Therefore, The purpose Flow Sequentializer is to generate flow sequence by steps of Flow Sorting, Equal Frequency Binning, and Vertical Flows Assembling. As a result, the flow in the same position will have the same characteristics, making the pattern of the flow sequence easy to learn.

\textbf{Flow Tokenizer.} For inter-flow correlation mining, it treats network flow as a new modality and aligns them to the semantic space of LLMs. It designs a learnable Flow Tokenizer which encodes flow sequences into a high-dimension token embedding. And inter-flow correlations will be explored by LLM. The embedding representations will be classified by a MLP, and output the binary label of malicious or benign.

\subsection{Evaluation}
The system is implemented on a Linux server with an AMD Ryzen-9 7950X 16-core CPU and an NVIDIA GeForce RTX-4090 GPU, and uses \textit{Llama2-7b-chat} as the backbone, configuring the model’s floating-point precision to bfloat16. 

DoLLM is evaluated in both simulation and real-world trace. In the simulation test, we mix the CIC-DDoS2019 and MAWI datasets to create two datasets, each with a 1:1 ratio of malicious to benign flows. The first dataset includes three attack vectors: DNS, NTP, and SYN. The second dataset consists of the remaining eight vectors from CIC-DDoS2019. We train on the former and test on the latter to evaluate zero-shot capability. In the real-world test, a NetFlow dataset is collected from a Top-3 nationwide operator, and the ratio of malicious to benign flows is 1:10. We compare the performance of DoLLM with four ML-based methods, including the supervised learning method XGBoost, the unsupervised anomaly detection method AutoEncoder, and two sequence-based models: LSTM and Transformer. These methods encompass the majority of current data-driven detection mechanisms. We use the F1-score as the evaluation metric.

\begin{figure}
    \centering
    \includegraphics[width=1\linewidth]{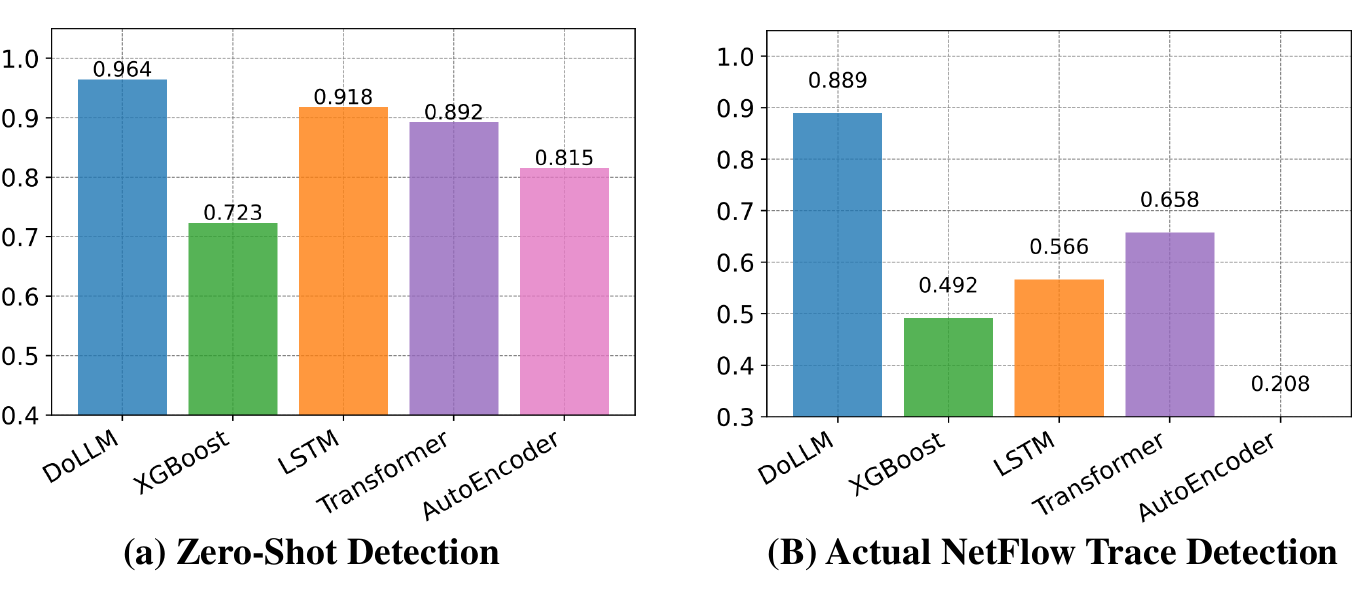}
    \caption{The F1 scores of various methods in zero-shot detection and real-world NetFlow trace.}
    \label{fig:dollm_f1}
\end{figure}

It shows the best detection performance across all scenarios. Especially, in zero-shot scenarios, its F1 score improved by up to $24\%$ compared to other methods, indicating its potential to detect new types of attacks. In real-world NetFlow trace, it improved by at least $35.1\%$ compared to other methods, showcasing its significant advantage in real-world scenarios. The test results for the above two scenarios are shown in Fig.~\ref{fig:dollm_f1}.

\section{Conclusion}
\label{Sec-conclu}
With the explosively break-down of LLM, Applications of LLM in network security has developed rapidly in recent years. This paper studies LLM-powered malicious traffic detection, from a novel perspective of what roles LLMs can function. We present a holistic view of the architecture of LLM-powered malicious traffic detection, including unsupervised pre-training, supervised fine-tuning and ICL, inference and detection. Research opportunities and challenges are next discussed. 
Finally, we present our case study on LLM-based carpet bombing DDoS detection, showcasing the innovative design. The proposed framework successfully achieve good accuracy for malicious traffic detection, surpassing existing system with a nearly 35\% improvement. 

The proposed LLMs architecture is a general framework for classification tasks. It can be not only applied to malicious traffic detection, but also to any other classification task. LLMs can act as the Classifier, Encoder, and Predictor to mine massive data, uncover the latent features and representations, and achieve the accuracy classification. With the remarkable capabilities of context-understanding and commonsense knowledge, LLMs would provide a new paradigm for traffic classification and network threat detections. In the future, we would continue exploring the potentials of LLMs in more network tasks, such as abnormal log analysis, malware detection, encrypted traffic classification, and beyond.


 
%
\bibliographystyle{IEEEtran}
\bibliography{main.bib}

\end{document}